\documentclass[twocolumn]{emulateapj}
\submitted{Science with Parkes @ 50 Years Young, 31 Oct. -- 4 Nov., 2011}

\shorttitle{The first search for glycine}
\shortauthors{Storey}

\begin{document}

\title{The first search for glycine and other biomolecules}

\author{J.W.V. Storey}
\affil{School of Physics, University of New South Wales, Sydney NSW 2052, Australia}
    
\email{j.storey@unsw.edu.au}

\begin{abstract}
In the 1970s the microwave spectroscopy group at Monash University became the first in the world to determine the spectral frequencies of urea, glycine, and  several other biomolecules.  We immediately searched for these at Parkes, using existing centimetre-wave receivers plus newly built receivers that operated at frequencies as high as 75GHz (and used just the central 17 m of the dish).  Although these searches were largely unsuccessful, they helped launch the now flourishing field of astrobiology.
\end{abstract}

\keywords{interstellar molecules, biomolecules, astrobiology}

\section{Introduction}

The 1970s were an exciting time in molecular-line radioastronomy.  The discoveries of ammonia and water in the interstellar medium with the Hat Creek antenna \citep{che68,che69} were followed by a flurry of discovery of new species---some expected  and some rather surprising.  The race was on to discover the largest, weirdest, and most significant new molecules.  In some cases, the searches were motivated by a desire to find answers to the most obvious of questions: for example, how on earth do these molecules form in the first place?  However, because the discovery space was so broad, the field so new and the level of ignorance near total, the motivation was sometimes more basic:  this molecule looks pretty neat---I wonder if we can detect it?

In many ways the excitement generated was similar to that currently experienced in the exoplanet community.  The existence of interstellar molecules, in regions as diverse as the Galactic Centre and the Orion gas clouds, was a clear indication that chemistry---and in particular organic chemistry---was alive and well across the galaxy.  Then, discoveries were made in the Magellanic Clouds and beyond. Organic molecules, the building blocks of life, were clearly distributed right throughout the universe.

This was a game anyone could play.  You needed access to a decent radio telescope, a talented engineering team who could tune the receiver (or build new ones) to get to frequencies not previously available, and of course a knowledge of what those frequencies were---preferably to at least six significant figures.  In some cases the particular transition frequencies had been directly measured in the laboratory and were already published in the open literature.  In some cases, the frequencies could be calculated with sufficient accuracy from the published rotational constants.  Directors of radio telescopes around the world were flooded with observing requests for all plausible molecules, and quite a few implausible ones as well: all that mattered was that you knew the frequencies.  An orgy of discovery followed.

Then it got interesting.  First, there was some weird stuff out there.  For example, hydrogen cyanide ($HCN$) might reasonably be expected.  It's small, made up of some of the most cosmically abundant elements, and because it's a linear molecule with a respectable dipole moment it has just a few, strong spectral lines---making it easy to detect.  But  {\em hydrogen isocyanide} ($HNC$)? Who expected that?  It doesn't even exist on earth. It is less stable than $HCN$, so if it were to accidentally form under terrestrial conditions, it quickly realises its mistake and adopts the lower-energy arrangement of the three atoms.  In space, however, its lifetime can be very long as it hangs around waiting for a collision partner willing to be complicit in its internal rearranging.

Second, astronomers had run out plausible species to look for, for which the transition frequencies were known.  Conventional wisdom on what new species might be detectable was not always respected by nature.  The way forward was thus for the established laboratory microwave spectroscopy groups to form an alliance with radio astronomers.  New species (or new transition frequencies of known species) could be measured in the laboratory, and then, clutching a list of frequencies, the observing team would descend upon whatever radio telescope they could get their hands on\footnote{Every few years, DVCs and academic managers decide that researchers need to be forced out of their ``silos'' to ``look across the fences'' and be coerced into becoming more ``interdisciplinary''.  In my experience, this is so unnecessary as to be completely nuts.  Given the whiff of a new idea, good researchers will leap the fence like the drover's dog on fair day and join in whatever new fun is around. }.

And so it came to be that the Microwave Spectroscopy Group in the Chemistry Department at Monash University rose to international prominence in this field.  Led by the late Professor Ron Brown, and ably assisted by Dr Peter Godfrey and Jon Crofts, the group promptly used Parkes to make the first discoveries of both methanimine \citep{god73} and thioformaldehyde \citep{sin73}. With one of the best laboratories in the world, the Monash group were well positioned to take on the multi-facetted challenge of positing plausible new chemical species, figuring out how to synthesise them if they didn't already exist, measuring with great precision their transition frequencies, then undertaking the necessary astronomical searches. In 1973 I joined this group as a PhD student.

\section{The dawn of astrobiology}

An obvious question to ask in 1973 was: how far along the route towards biologically important molecules does synthesis in the interstellar medium occur?  Some 30 different interstellar species were known, including methanol, formamide and some cyanoacetylenes.  Furthermore, amino acids had been found in the Murchison meteorite \citep{kve70}, and these had presumably formed well away from the earth.

Unfortunately, none of the amino acids had a known microwave spectrum.  Indeed, the conventional wisdom was that upon heating, even the simplest amino acid (glycine) decomposes.  If true, this would be a major setback---the microwave transitions, which couple different energy levels of the freely rotating molecule, can only be properly observed in the gas phase.

So, while discovering interstellar glycine would be the glittering prize, we decided to take things a bit more slowly and tackle something a bit easier: urea.

\section{Urea}

Urea, $CO(NH_2)_2$, is a quintessentially ``biological'' molecule with an interesting history.  It was first discovered by Hilaire Rouelle in 1773 by evaporating dog urine.  While this might seem a curious pastime, these were very early days for chemistry: only three years previously had Priestley  discovered oxygen.  The manner of urea's discovery urea saw it categorised at the time by the ``Vitalists'' as something mystical: something that it is a part or product of living organisms, rather than something like iron that can be made by man out of rocks.

In 1828, W\"{o}hler threw the cat amongst the pigeons by synthesising urea from inorganic materials.  Suddenly there was nothing special about biological material, and the Vitalists were on a slippery slope to oblivion.

Nowadays, urea is manufactured in industrial quantities for plastics (eg., urea formaldehyde foam), as fertiliser, as a flavour enhancer for cigarettes, and is handy for putting together improvised explosive devices.  It melts at 134$^\circ$C.  In principle, a length of microwave waveguide heated to appropriate temperatures and containing a small sample of urea might be expected to allow observation of its spectral lines.  However, early indications were that urea simply falls apart instead of vaporising---as evidenced by the observation of enormously strong ammonia lines in the spectrometer.  But perhaps, if conditions were just right, enough urea vapour might be produced to allow its spectrum to be measured.

This motivated the construction of the Heated Waveguide Cell (Fig 1.)  It proved a huge success; spectral lines of urea were not only seen but assigned to their corresponding transitions \citep{bro75}, and an interstellar search was at last possible.  Thus, in 1974, Parkes undertook the first ever search for interstellar urea (Fig. 2).  While we did manage to set the PDP9 computer on fire (Fig. 3), no urea was detected.

\begin{figure*}
\centerline{\includegraphics[width=150mm]{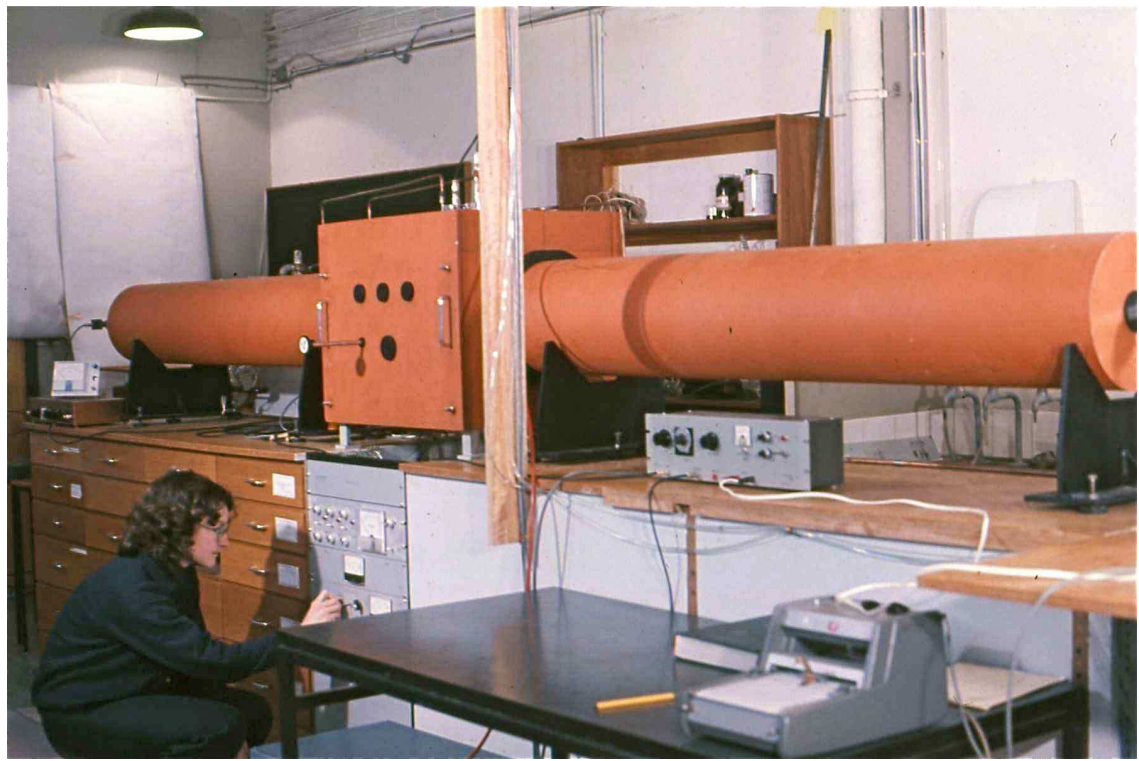}}
\caption{Marie-Paule Bassez and the Heated Waveguide Cell, 1976.}\label{fig1}
\end{figure*}

\begin{figure*}
\centerline{\includegraphics[width=150mm]{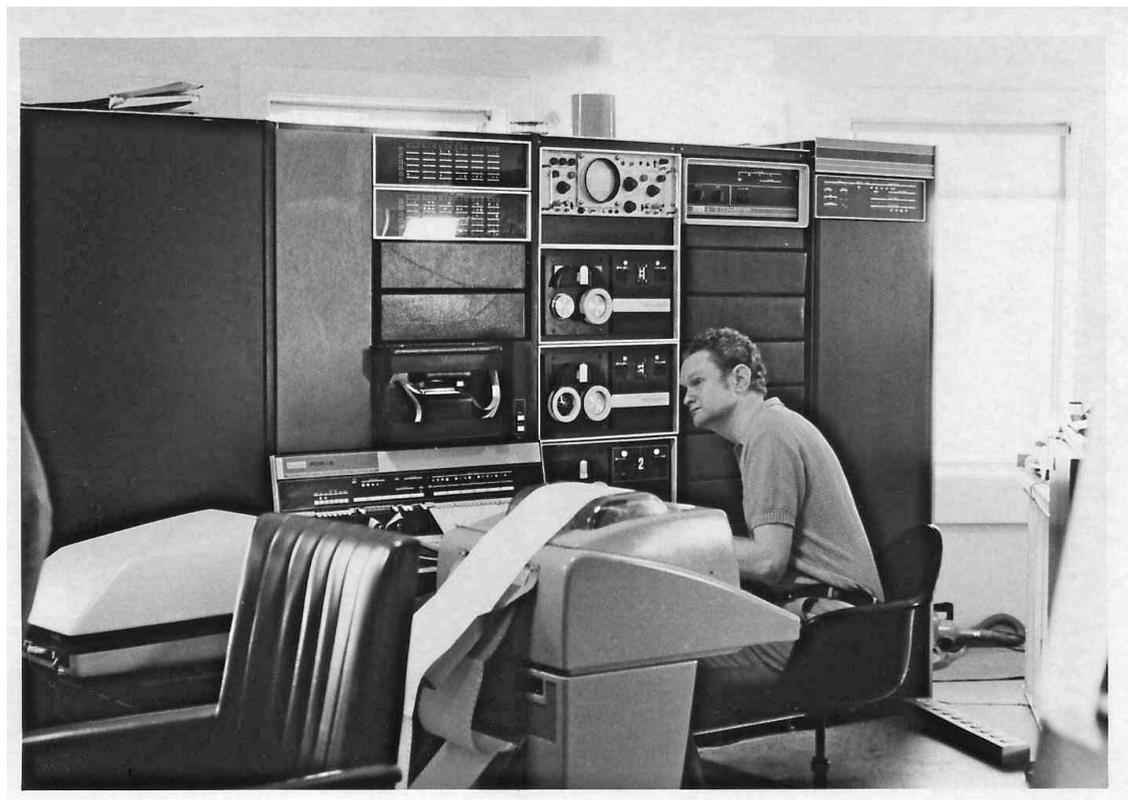}}
\caption{Peter Godfrey at the PDP9 in the Parkes control room, 1974.}\label{fig2}
\end{figure*}

\begin{figure*}
\centerline{\includegraphics[width=150mm]{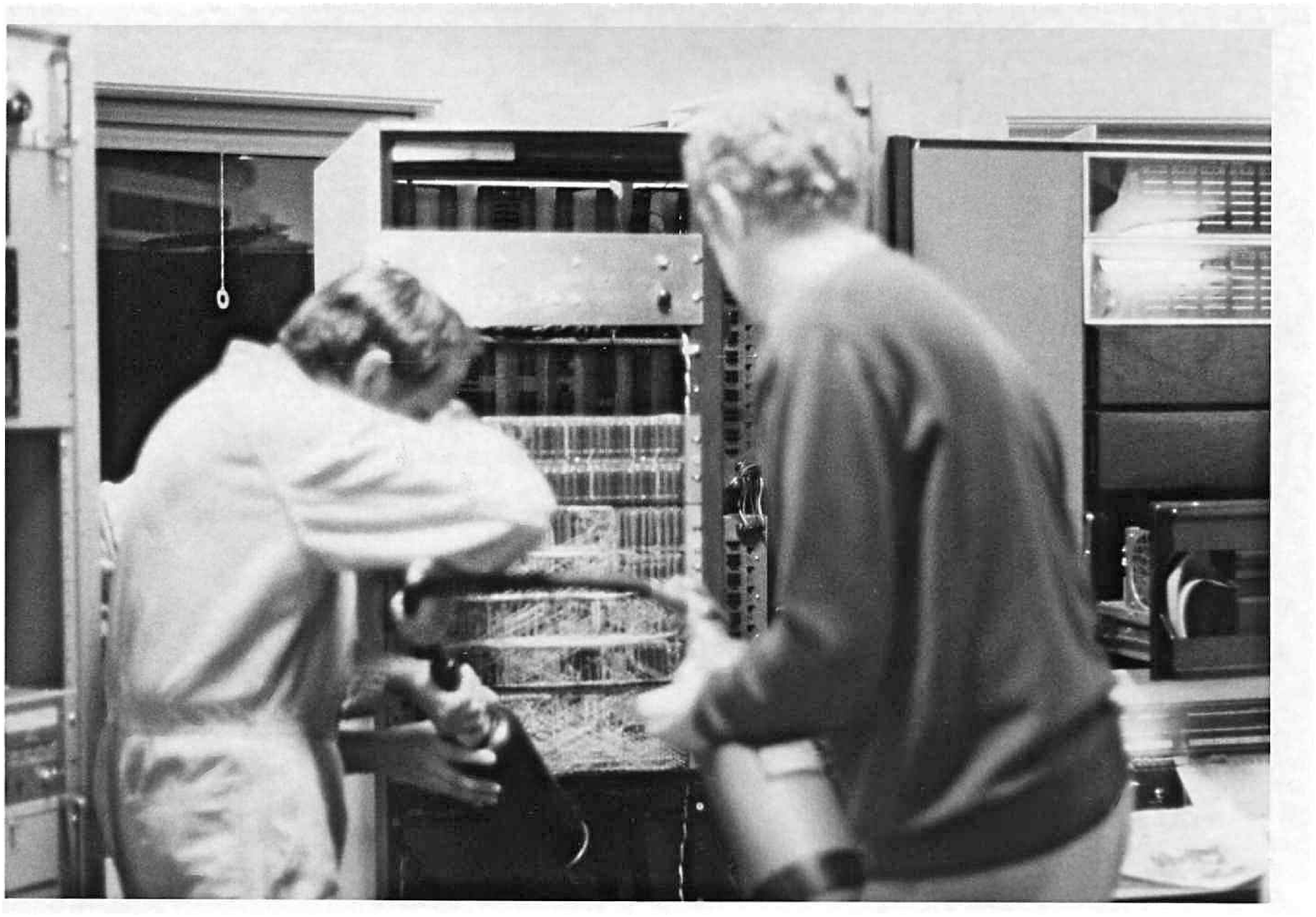}}
\caption{Brian Robinson and Peter Godfrey extinguish the PDP9 fire, 1974.}\label{fig3}
\end{figure*}

A further search was conducted with what was then the 36 foot telescope at Kitt Peak in Arizona.  Again, no urea was found but, in keeping with the pace of discovery of the time, we confirmed the identity of the line attributed to $HNC$ by detecting the rarer $HN^{13}C$ \citep{bro76}.

Thirty-four years later, \citet{kuo10} have just reported the first detection of interstellar urea, using CARMA. 

\section{Aminoacetonitrile}

Though not technically a ``biological'' molecule, aminoacetonitrile, $NH_2CH_2CN$, is big enough and floppy enough to be one, and is plausibly a precursor to glycine.  We did some laboratory work to confirm the frequencies \citep{bro77}, then searched (unsuccessfully) at Parkes.  It is  sobering to see now how far away we were in 1974 from any possibility of success: aminoacetonitrile was not detected until 2008 \citep{bel08}, and then it required the combined resources of the IRAM 30 m, IRAM interferometer, and ATCA.  Belloche et al are gracious in their omission of phrases such as ``naive optimism'' from their report of our earlier attempts.

\section{Glycine---at last}

Glycine, $NH_2CH_2COOH$, is the simplest amino acid.  It was discovered by Henri Braconnot in 1820, and first synthesised in 1858.  It is the only non-chiral member of the twenty ``standard'' amino acids. Normally existing as a crystalline solid, or as an important component of some living creature (it makes up some 45\% of silk protein, for example), it enters the gas phase reluctantly---so reluctantly that all attempts to make it do so had hitherto been unsuccessful.  At Monash, the Heated Waveguide Cell that had proved so useful for urea yielded not a single line we could attribute to glycine.

Clearly a new approach was needed, and so the Heated Parallel Plate Cell was born (Fig. 4).  Space does not permit a detailed description (which can be found elsewhere, \citet{sto76}) of this grossly over-engineered 230 kg piece of hardware, but the basic idea was to vapourise the glycine and pass it rapidly {\em across} a parallel-plate microwave transmission line, causing the molecules to interact with the microwaves before they had a chance to either decompose or turn into some less interesting.

\begin{figure*}
\centerline{\includegraphics[width=150mm]{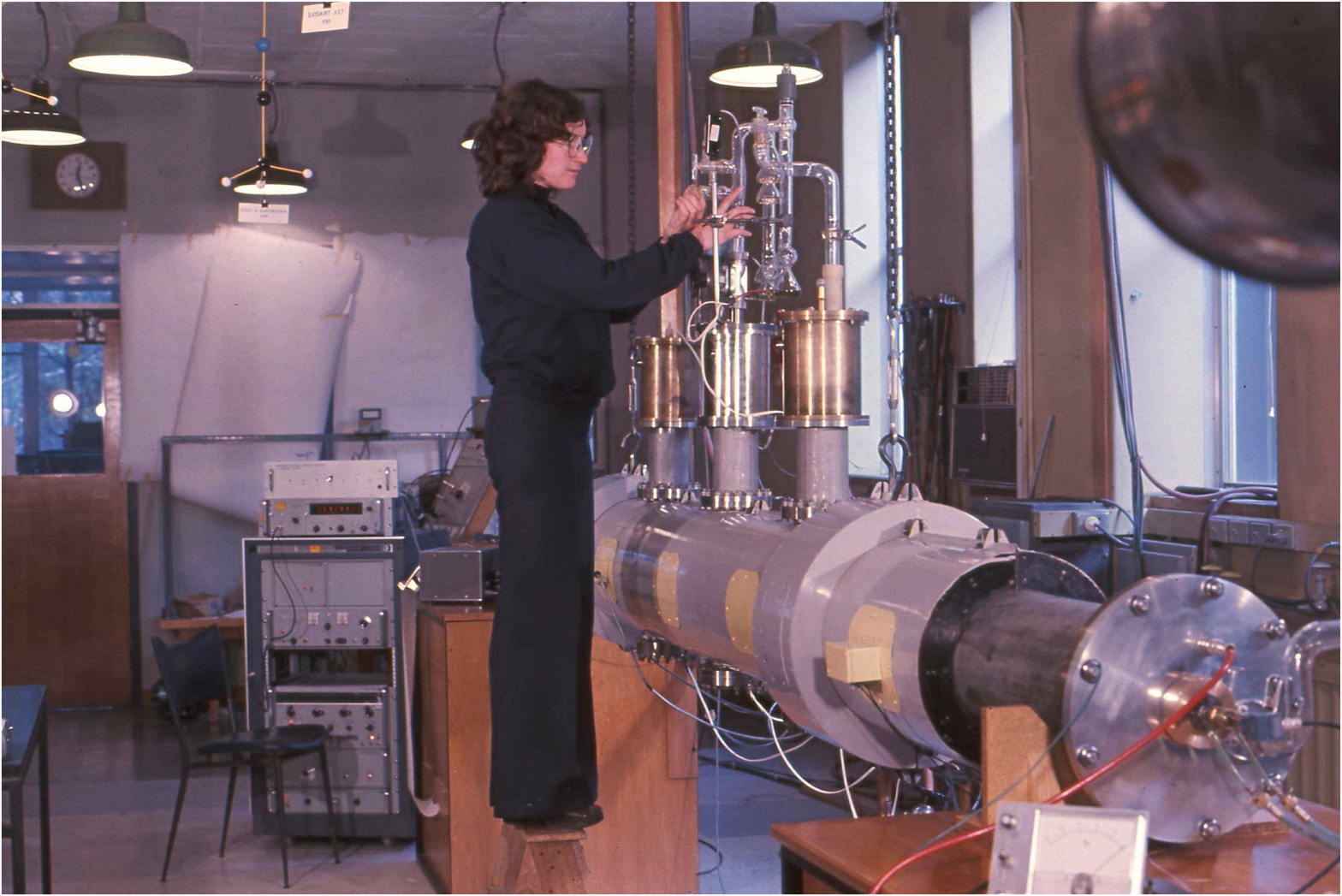}}
\caption{Marie-Paule Bassez and the Heated Parallel Plate Cell, 1976.}\label{fig4}
\end{figure*}

Fortunately, this new approach worked---glycine lines were identified \citep{bro78}, assigned to specific transitions (allowing a molecular structure to be derived), and at last offering the possibility of conducting an interstellar search.  The Monash group were quick to do exactly that \citep{bro79}. As well as using the  Kitt Peak and Onsala dishes, observations were made at Parkes at the outrageous frequency of 75GHz using a cooled mixer and a newly installed set of precise panels on the central 17 m of the dish.

Thus began the long series of searches, by many groups across the world, that have made glycine the most unsuccessfully searched-for interstellar molecule of all time.  \citet{sue78} identified a second, lower-energy  conformer of glycine in laboratory spectra, but searches for that, starting with \citet{hol80}, have been equally unproductive.  In 2003 \citet{kua03} reported a tentative detection, but this was first disputed by \citet{sny05} before being finally discredited by \citet{jon07}. As of March 2012 no definitive detection of any glycine conformer has been reported.   The last word perhaps belongs to \citet{lat11} who show that, like $HNC$, glycine is not the lowest energy arrangement of those particular atoms.  They go on to state: {\em``...our conclusion is that this search will remain extremely difficult with the present instruments and we propose searching instead for other examples among the most stable isomers.''} [ie, forget glycine; go for something else instead].

\section{The future}
To Lattelais et al one might respond: ``{\em But what about the case of $HNC$?}"  Nevertheless, it does appear that neither glycine, nor any similar biological molecule, is detectable with current technology.  Some 36 years after the first laboratory measurements of its spectral lines, glycine remains as elusive as ever.  Has too much time already been spent searching for it, and indeed what would be the real significance of its discovery?  Whatever the rational view might be, finding an amino acid in space has such a popular appeal that the search will no doubt continue.  I wish those researchers the best of luck.

\acknowledgments

I thank all those people who helped make my PhD years so enjoyable. Dr Peter Godfrey was a wonderful mentor to me during this time.  Special mention must also be made of the late Brian Robinson and the late Ron Brown, both of whom were exceptional in their respective fields.    Particular thanks, too, to Mike Ballister, Marie-Paule Bassez, Bob Batchelor, Jon Crofts, Gerry McCulloch and Mal Sinclair, and to the many students  postdocs and workshop staff at Monash with whom I worked.

\end{document}